\documentclass[twocolumn]{aastex631}
\usepackage{amsmath}

\newcommand{\gr}{$\gamma$-ray}

\shorttitle{An optical QPO in TXS 1206+549}
\shortauthors{Ren Sun \& Zhang}
\graphicspath{{./}{figures/}}

\begin{document}

\title{A Possible Optical Quasi-periodic Oscillation of 134~day in the Radio-loud Narrow-line Seyfert 1 Galaxy TXS 1206+549 at $z=1.34$}

\author{Chongyang Ren}
\affiliation{Department of Astronomy, School of Physics and Astronomy, Key Laboratory of Astroparticle Physics of Yunnan Province, Yunnan University, Kunming 650091, People's Republic of China; zhangpengfei@ynu.edu.cn}

\author{Sisi Sun}
\affiliation{Department of Astronomy, School of Physics and Astronomy, Key Laboratory of Astroparticle Physics of Yunnan Province, Yunnan University, Kunming 650091, People's Republic of China; zhangpengfei@ynu.edu.cn}

\author{Pengfei Zhang}
\affiliation{Department of Astronomy, School of Physics and Astronomy, Key Laboratory of Astroparticle Physics of Yunnan Province, Yunnan University, Kunming 650091, People's Republic of China; zhangpengfei@ynu.edu.cn}

 \begin{abstract}

Here we report an optical quasi-periodic oscillation (QPO) with a period of $\sim$134~day detected
in g- and r-band light curves of a narrow-line Seyfert 1 galaxy TXS 1206+549 at redshift of 1.34
with the data from the Zwicky Transient Facility (ZTF) observations.
After considering the trial factor, the significance levels in the two bands are 3.1 $\sigma$
and 2.6 $\sigma$, respectively.
The QPO signal presents about 10 cycles ranging from 2018 March to 2021 December lasting $\sim$4 year.
A near-sinusoidal profile also appears in the folded light curves by a phase-resolved analysis.
Interestingly, in the simultaneous light curve as the time scale of ZTF observations,
a potential periodic signal with similar period is detected in the o-band light curve from
the Asteroid Terrestrial-Impact Last Alert System data, additionally a weak peak is also detected
at the similar period in the gamma-ray light curve obtained from the \emph{Fermi} Gamma-ray Space Telescope data.
Some potential origins of periodicities in active galactic nuclei are discussed for
the QPO reported here.

\end{abstract}
\keywords{Active galactic nuclei (16); Time series analysis (1916); Period search (1955)}

\section{Introduction}
\label{Intro}

Active galactic nuclei (AGNs), hosting a supermassive black hole (SMBH) over $10^8$ M$_\odot$,
are the most luminous sources of electromagnetic radiation in the universe,
the majority of physical processes are launched from their large-scale powerful relativistic jets.
AGN multi-wavelength electromagnetic spectral energy distribution (SED) extends
from MHz radio frequencies to TeV gamma-ray energies \citep{umu97}.
And the Narrow-line Seyfert 1 (NLSy1) galaxies are a unique subclass of AGNs \citep{op85},
they have some properties including of narrow width of the broad Balmer emission lines and weak forbidden lines \citep{g89},
strong permitted optical/UV Fe$_{\rm II}$ lines \citep{so81}, soft X-ray excess \citep{ojh+20},
near-Eddington accretion rates \citep{min+00,rak+17}, and smaller black hole (BH) masses.
Their amplitudes of optical light curves are usually lower comparing with other broad-line AGNs \citep{rs17},
and the variabilities may be resulted in the dynamical processes close to BHs,
which play an important role in revealing AGN radiation mechanisms and structures.

Recently many interesting quasi-periodic oscillation (QPO) cases are detected
in the electromagnetic emissions of AGNs at multiwavelength from radio frequencies to GeV $\gamma$-rays,
the scale of periods of the QPO signals span the timescale ranges from minutes
to years \citep[][and references therein]{khm+13,b17,lly+17,b18,rzz+21,zw21,kts92,
fl00,vls+06,vln+08,zzw+14,sct14,vzg+23,gmw+08,lig+13,pyy+16,zzy+17,zzl+18,1553+15,sct16,
zyl+17,zyz+17,zzz+17,yzz+18,zwc+18,zyz+20,yyz+21,zyz+21,gzy+22,cyg+22,gtz+23}.
Several special cases of them are summarized here, for example,
\citet{zw21} claimed a discovery of a 176~day QPO of the NLSy1 galaxy J0849+5108
in the 11~yr radio light curve obtained with the Owens Valley Radio Observatory
40~m telescope at 15~GHz. 
Up to now, it is the longest duration QPO signal found in AGNs, has been presented for 21 cycles.
Recently, \citet{zwg+22} reported double yearly QPOs
of a Seyfert type galaxy NGC 1275 (or 3C 84 in radio) in 1.3 mm wavelength light curve monitored by
the Submillimeter Array. Moreover, even in the optical polarized emission
of the blazar sources, the QPO signals are reported in blazars B2 1633+38 and
PKS 1222+216 with periods of years \citep{oab+20,zw22}.
Due to the current limited understanding of the variability characteristics of AGNs,
there is a considerable debate about the real statistical significance of these QPO findings.
A huge sample of the long-term light curves for AGNs have been provided by
the Zwicky Transient Facility (ZTF) observations over four years with
measurements made every few days \citep{ztf1+19,ztf2+19}, which provide us
a great convenience to study QPO signals in the long baseline
time series. Indeed some QPO cases have been detected in \citet{z22,bnj+22}
from the ZTF survey.

TXS 1206+549 is a radio and $\gamma$-ray loud NLSy1 galaxy and has a flat radio spectrum \citep{rss+21}.
As the blazars and other $\gamma$-ray NLSy1 galaxies, their broad-band SED shows the typical two-bump structure.
They exhibit strong variability in the universal electromagnetic emissions in the optical,
infrared, and $\gamma$-ray bands.
Up to now, TXS 1206+549 is the most distant NLSy1 galaxy at $z=$1.344 having $\gamma$-ray emissions \citep{rss+21},
while other (only about a dozen) $\gamma$-ray NLSy1 galaxies were detected at $z<$1.
Recently, we conducted a study on the phenomena of $\gamma$-ray eruptions exhibiting higher flux
and shorter duration in \emph{Fermi}-LAT observations, then we focused on the target TXS 1206+549.
During the multi-wavelength studies on this source,
we accidentally discovered that its optical light curves may exhibit periodic behavior.
Subsequently, we carried out a timing analysis for the $\sim$4 year optical light curve of TXS 1206+549
with the g- and r-band data from ZTF survey.
And a QPO signal was detected at both bands having a similar period of $\sim$134 day presenting about 10 cycles.
After considering the trial factor, the significance levels in the two bands are 3.1 $\sigma$
and 2.6 $\sigma$, respectively.
Interestingly, in the simultaneous light curve as the time scale of ZTF,
a potential QPO signal with a similar period was also found in the o-band light curve from
the Asteroid Terrestrial-impact Last Alert System \citep[ATLAS;][]{allas18}.
TXS 1206+549’s $\gamma$-ray emissions have been detected by the Large Area Telescope (LAT) onboard
the \emph{Fermi} Gamma-ray Space Telescope \citep[\emph{Fermi}-LAT;][]{Atwood2009} since 2008.
And a weak signal was also detected in the nearly simultaneous light curve as ZTF observations
at a similar period in 0.1--500.0 GeV.

We describe the data analysis of the ZTF optical light curve, ATLAS o-band light curve,
\emph{Fermi}-LAT $\gamma$-ray data, and the main results in Section~\ref{sec:data}.
Results for periodic analysis are shown in Section~\ref{sec:qpo}.
Summary and discussion are shown in Section~\ref{sec:dis}.
In the following text, numbers in parentheses represent the corresponding uncertainties on the last digit.

\section{Data Preparation and Reduction}
\label{sec:data}

\subsection{Light curves from ZTF}
\label{sec:ztf}
Zwicky Transient Facility is a new 48 inch Samuel Oschin Telescope at Palomar Observatory with a 47 square degree
field of view camera in optical wavelength \citep{ztf1+19,ztf2+19}.
It has scanned the entire Northern sky every two days started in 2018,
and released high quality data products every two months.
Such an observation cadence makes ZTF archival data very suitable for time-domain science,
especially for the exploring of periodic variabilities in the timescales of sub-year-long ones.
The ZTF released data can be obtained from the website of
the archive science data\footnote{https://irsa.ipac.caltech.edu/Missions/ztf.html}.

\begin{figure*}
\centering
\includegraphics[angle=0,scale=0.6]{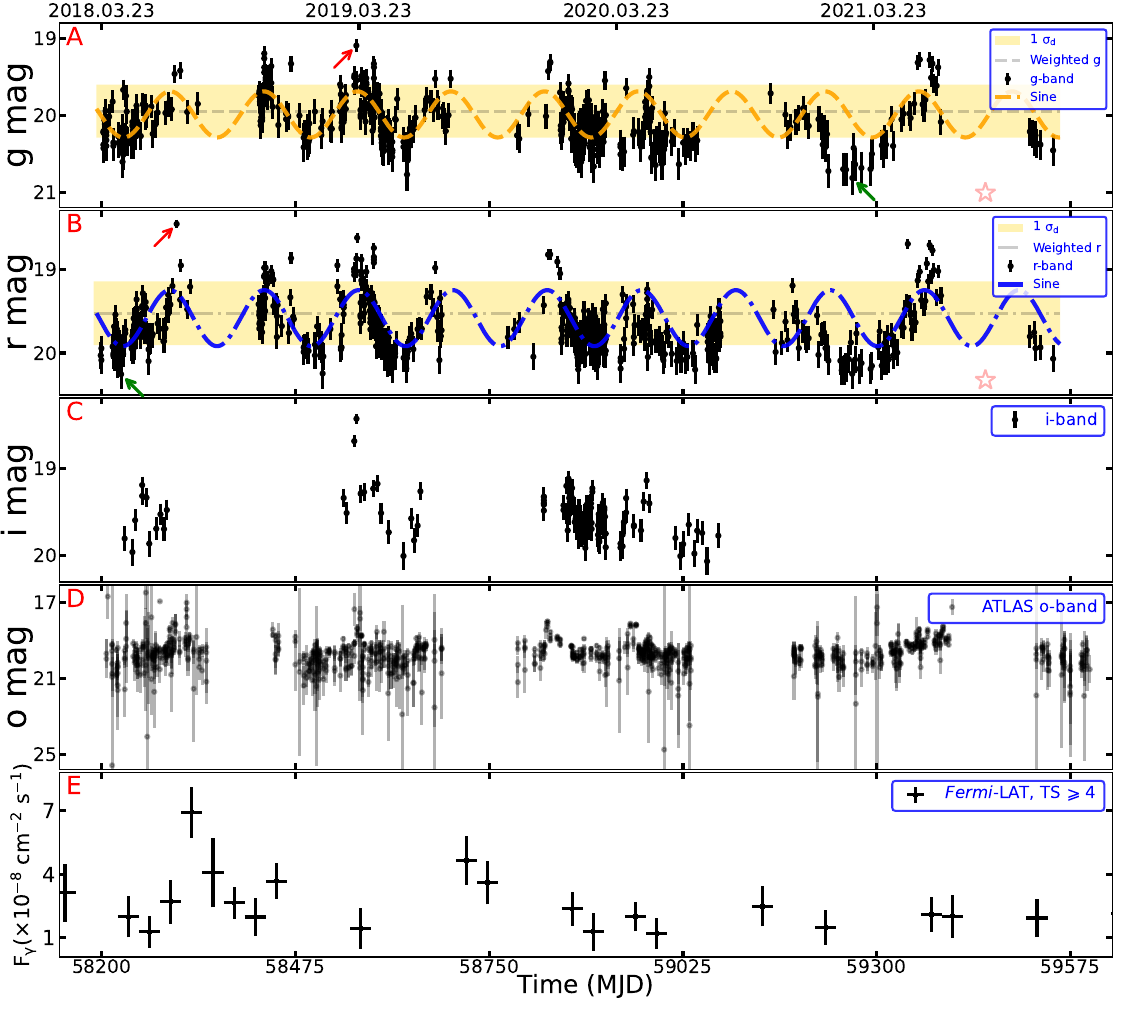}
\caption{Optical and $\gamma$-ray brightness measurements for TXS 1206+549.
        Optical g-, r-, and i-band light curves monitored by the ZTF observations
        are shown in panels A, B, and C, respectively. The orange dashed
        and blue dashed-dotted curves (in panels A and B) stand for the sinusoidal
        models in g- and r-bands, respectively, their parameters were
        given by the LSP code. 
        O-band light curves from the ATLAS observations is shown in panel D.
        The nearly simultaneous 0.1--500 GeV light curves with time-bin of 30 day with
        observations by the \emph{Fermi}–LAT shown in panel E.}
\label{fig:lc}
\end{figure*}

The source TXS 1206+549 (at the same coordinate as it in 4FGL)
is one of ZTF targets. It has been monitored more than
1300 day spanning from 2018 March 23 to to 2021 December 2 (MJD 58,200-59,550) in the g-, r-, and i-bands.
We show the optical light curves of g-, r-, and i-bands by ZTF in the panels A, B,
and C of Figure~\ref{fig:lc}, respectively.
The median and minimum time intervals of two adjacent data points of g-band (r-band)
are $\sim0.927$ (0.829) and 0.001 (0.001) day, respectively, the maximum interval is 126.266
(126.262) day at MJD 59,454 (marked with a pink star in Figure~\ref{fig:lc}).
For g-band, the brightest and faintest epochs of target are at MJD 58,561.3 and 59,265.4
with the magnitudes of 19.1 (indicated with a red arrow) and 20.8 (a green arrow), respectively.
The weighted magnitude was calculated and weighted by the errors of data points with
a formula of $\overline{M}=\sum{w_i*M_i}$, where $w_i=\frac{1}{\sigma^2_i}\frac{1}{W}$,
$W=\sum{\frac{1}{\sigma^2_i}}$, and $\sigma_i$ indicating the errors.
And the weighted standard deviation ($\rm\sigma_d$) was calculated by
$\rm\sigma_d$=$\sum{w_i*(M_i-\overline{M})^2}$, its value is 34.4\%.
For r-band, the corresponding epochs are MJD 58,306.2 and 58,227.3
at the magnitudes of 18.5 and 20.3 shown with red and green arrows in B panel of Figure~\ref{fig:lc}.
Its values of $\overline{M}$ and $\rm\sigma_d$ are 19.53 and 38.0\%.
We show that weighted results of g- and r-bands as horizontal gray lines and yellow shaded regions, respectively.

\subsection{Light curve from ATLAS}

Asteroid Terrestrial-impact Last Alert System\footnote{https://fallingstar-data.com/forcedphot/}
consists of four 0.5~m telescopes with two of them in Hawaii, one each in Chile and South Africa \citep{allas18},
which provides public access to photometric measurements over its full history survey.
TXS 1206+549 is one of the objects monitored by ATLAS. We obtained its o-band light curve from
the website of archive data\footnote{https://fallingstar-data.com/forcedphot/queue/}.
ATLAS observations for TXS 1206+549 span from 2015 December 9 to 2023 March 28 (MJD 57,365.6-60,031.4),
we show the o-band light curve with a simultaneous observations of ZTF in D panel of Figure~\ref{fig:lc}.
For the simultaneous o-band light curve, the median, minimum and maximum time intervals of
two adjacent data points are 1.6, 0.0005 and 146.3~day.
Its minimum, maximum, and median magnitudes are $\sim$15.3, 25.6 and 19.7 respectively.
It should be noted that data quality of ATLAS light curve is poorer than that obtained by ZTF
(see D panel of Figure~\ref{fig:lc}), and
the ATLAS light curve is used only as an independent check for that QPO detected by ZTF observations.

\subsection{\texorpdfstring{\gr}{gamma-ray}~light curve from Fermi-LAT}
\label{sec:fermi}
The source TXS 1206$+$549 was cataloged in the fourth Fermi Large Area Telescope catalog \citep[4FGL-DR3,][]{4fgl-dr3}
and named as \gr~point source 4FGL J1208.9$+$5441 with a coordinate
(R.~A.~=~$\rm12^h08^m54^s.264$, decl.~=~$+54^{\circ}41^{'}58^{''}.208$).
In the \gr~data analysis, the Pass 8 $Front+Back$ SOURCE class events (i.e., evclass=128 and evtype=3) 
were selected with the energy range of 0.1–500 GeV spanning from 2008 August 4 to 2022 June 22 centered at target's position
within a $20^{\circ}\times20^{\circ}$ region of interest (RoI).
The events were reduced by a selection of the zenith angle $\leq90$ and an expression of
"DATA\_QUAL$>$0\&\&LAT\_CONFIG==1"
to obtain those high-quality events in the good time intervals.
Then a binned maximum likelihood analysis was performed between the whole data and a model file.
The model contains the parameters of the flux normalizations and spectral shapes for
all known 4FGL sources in the RoI. In 4FGL-DR3, the \gr~emissions from TXS 1206$+$549 was described by
a log-parabolic spectrum model with a form of $dN/dE=N_0(E/E_b)^{-\alpha-\beta \log(E/E_b)}$.
We obtained an average photon flux of $3.13(2)\times10^{-8}$~photons~cm$^{-2}$~s$^{-1}$
in the 0.1–500~GeV energy range with a test statistic (TS) value of 2,810.18.
Its best-fit spectral parameters $\alpha$, $\beta$, and $E_b$ are 2.53(1), 0.07(1), and 0.42(1) GeV, respectively.
Those values are in agreement with those reported in 4FGL-DR3.
All the corresponding best-fit parameters 
were saved as a new model file, and the following light curve analysis was carried out based on this model.

\begin{figure}
\centering
\includegraphics[angle=0,scale=0.7]{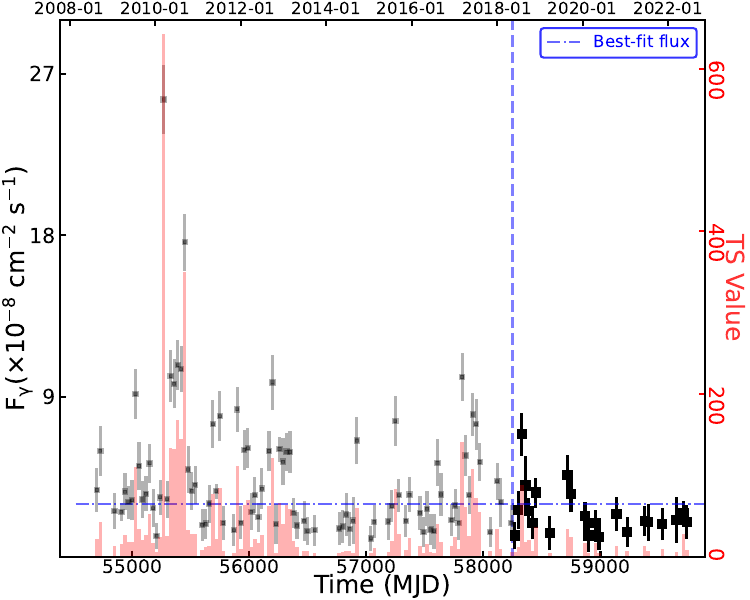}
\caption{TXS 1206+549 $\gamma$-ray light curve in 0.1–500 GeV with \emph{Fermi}–LAT
         during the whole observations, data points with TS $\geqslant$ 4 are selected.
         The blue dashed-dotted line represents the average integrated photon flux.
         The vertical dashed line indicates the starting epoch of ZTF observations.
         The corresponding TS values for data points are shown with pink bars.}
\label{fig:latlc}
\end{figure}

Then a monthly light curve with 0.1–500.0 GeV energy range was constructed by performed
an unbinned likelihood analysis during the corresponding observations. In this analysis,
we only freed the flux normalizations for the sources within $5^{\circ}$ RoI,
others were fixed at their best-fit values. In order to catch the variations as complete as possible,
the data points with TS $\geqslant$ 4 are included in the following periodic analysis.
We show flux variations and their corresponding TS values for each time-bin as black crosses and pink shaded regions in Figure~\ref{fig:latlc}, respectively.
And the nearly simultaneous light curve with ZTF observations is shown
in E panel of Figure~\ref{fig:lc} to facilitate comparison of multi-wavelength variabilities.
The average photon flux over the whole \emph{Fermi}-LAT observations in 0.1–500.0 GeV is shown as a blue 
dashed-dotted line.

\section{Periodic analysis and Results}
\label{sec:qpo}
\subsection{Periodicity Search}
Two different mathematical techniques, the generalized Lomb-Scargle periodogram
\citep[LSP;][]{l76,s82,zk09} and the weighted wavelet Z-transform \citep[WWZ;][]{f96},
were employed in the periodic variability analysis for optical and \gr~light curves.
These independent techniques were used to check each other results.
In the optical g- and r-bands data from ZTF observations,
we found a periodic signal with a sharp peak at the similar period in their power spectra.

\begin{figure*}
\centering
\includegraphics[angle=0,scale=0.49]{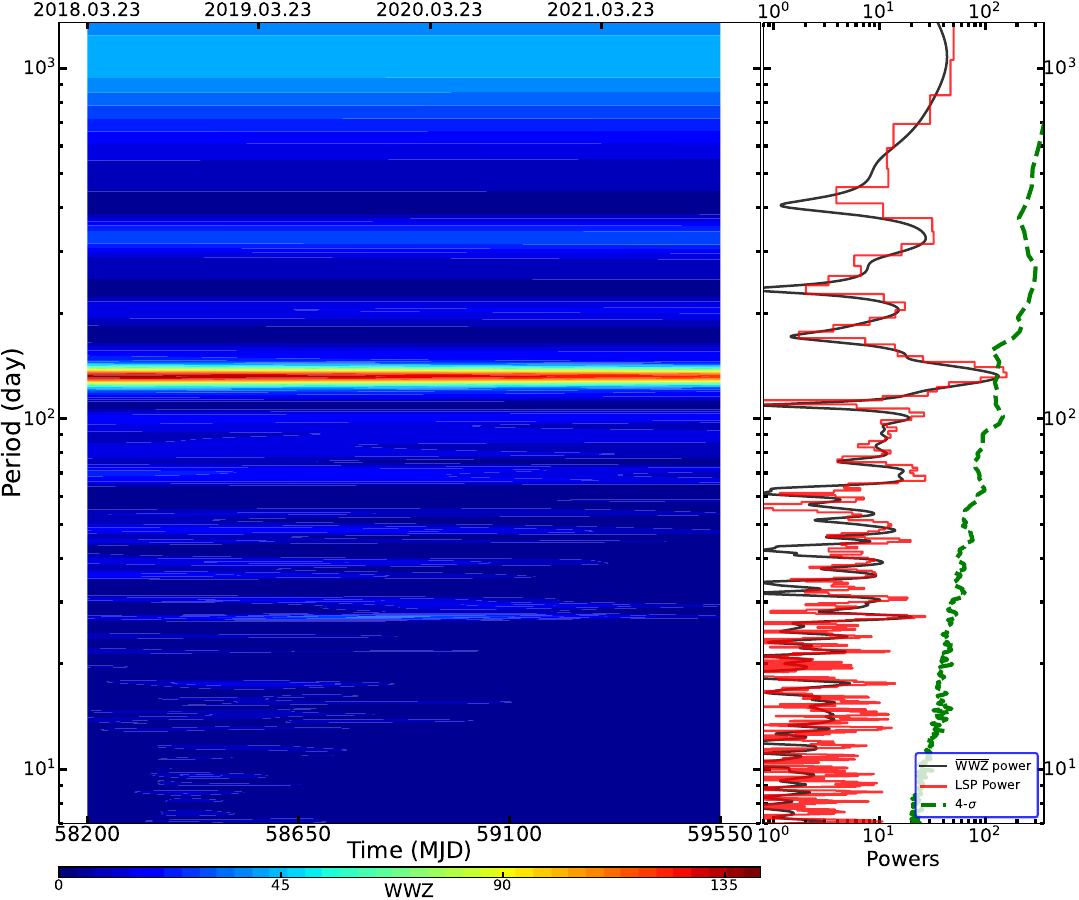}
\includegraphics[angle=0,scale=0.49]{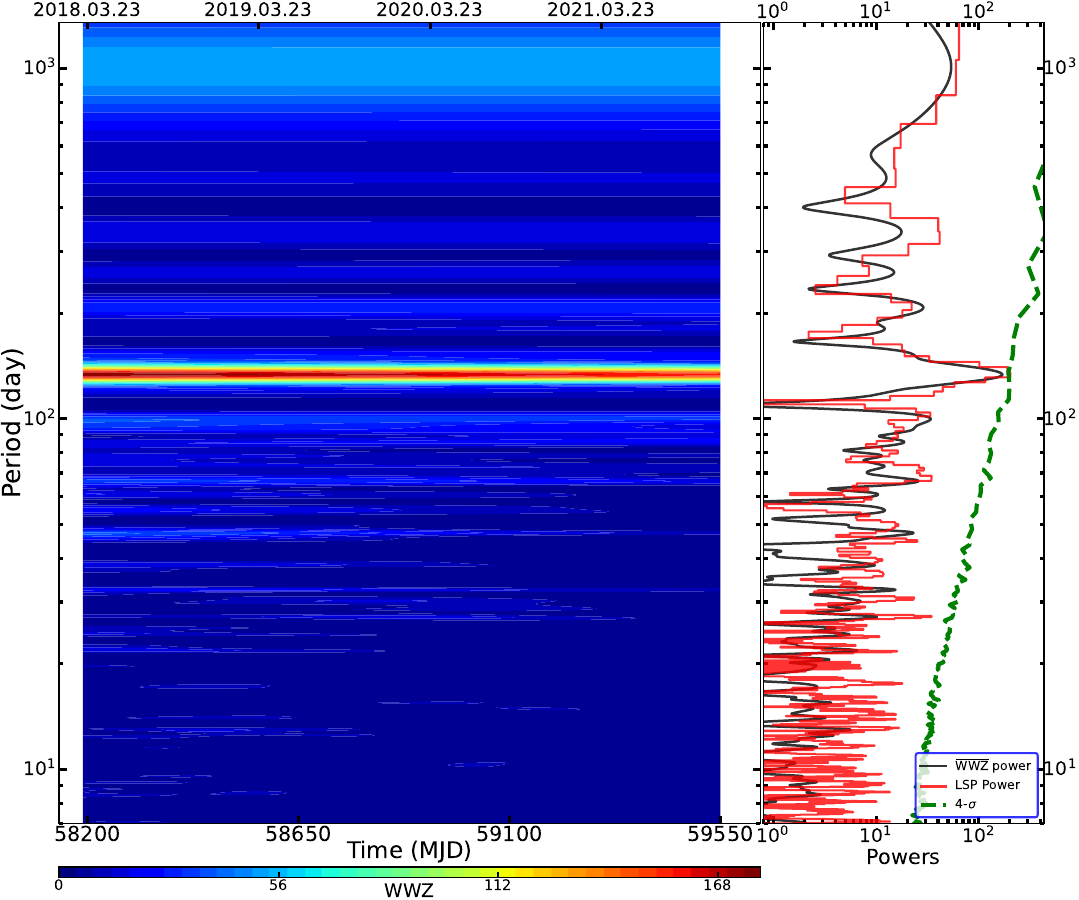}
\caption{TXS 1206+549 power spectra by constructed with the techniques of LSP and WWZ for
         the optical light curves of g- and r-bands covering the time range of MJD $\sim$58200–59550.
         The LSP, time-averaged WWZ, and WWZ power spectra are shown with red histogram, black curve,
         and color-scaled WWZ power, respectively. The spectra of g- and r-bands are shown in
         the left and right panels respectively.
         A obvious power peak is shown in the LSP and WWZ power spectra with a similar
         frequency corresponding to $\sim$134~day.
         The green dashed curves stand for 4$\sigma$ significance level derivd from the artificial light curves.}
\label{fig:gr-spec}
\end{figure*}

For the g-band light curve, we constructed its LSP power spectrum
and show it in left panel of Figure \ref{fig:gr-spec} with red histogram in log-log units.
From it, a significant QPO signal presents in the LSP power spectrum.
The periodical variations can be described by a sinusoidal function with a form of $A\sin([2\pi(t-t_0)/P])+A_0$.
Their parameters and uncertainties were derived by the LSP code, and their values are 0.300$\pm$0.016,
132.8$\pm$0.7~day, 58197.5$\pm$1.2, and 19.99$\pm$0.01 for $A$, $P$, $t_0$ and $A_0$, respectively.
We show the sine in A panel of Figure~\ref{fig:lc} with an orange dashed sine.
From it, about ten cycles of periodic modulation appears in the light curve.

\begin{table}
\begin{center}
\caption{PSD best-fit parameters}
\begin{tabular}{cccccc}
\hline\hline
& $K$ & $\alpha$ & $\beta$ & $f_{\rm bend}$ & $c$ \\
& $\times10^{-5}$& & & $\times10^{-2}$ day$^{-1}$ & $\times10^{-6}$\\
\hline
g-band & 1.88(41) & 1.00(17) & 2.47(22) & 4.63(147) & 1.02(40)\\
r-band & 2.72(93) &	1.00(14) & 2.39(33) & 2.15(107) & 9.45(354)\\
\hline
\end{tabular}
\label{tab:par}
\end{center}
{\bf Notes.} {Parameters of the underlying best-fit PSD obtained by the latest
implementation (https://github.com/lena-lin/emmanoulopoulos).}
\end{table}

A periodic signal or statistical fluctuations in a time-series may lead to
a power peak in a power spectrum. In order to determine the significance level
for periodicity signal, we employed a program for simulating light curve basing the power spectral
density (PSD) and probability density function (PDF) of variations derived from the original light curve,
as that provided by \citet{emp13}.
In general, the power spectra of AGN light curves are typically described
using a power-law (PL) or a smoothly bending power law plus a constant (SPL) models \citep{gv12}.
In our analysis, the SPL model yielded the better maximum likelihood fitting results than the PL one.
So, in the subsequent simulated light curve, we adopted the SPL model as the underlying best-fit PSD.
We used the SPL, $PSD(f)=Kf^{-\alpha}[1+(f/f_{\rm bend})^{\beta-\alpha}]^{-1} + c$,
to estimate the underlying PSD \citep{gv12}.
Then the maximum likelihood fitting was employed to obtain their best-fit parameters for the PSD \citep{bv12}.
In the fitting, we make the assumption that the components of the periodogram asymptotically follow
a $\chi^2$ distribution around the underlying PSD model.
Employing maximum likelihood estimation and minimizing the negative logarithmic likelihood function,
we obtained the best-fit parameters. We present the best-fit parameters along
with their associated relative uncertainties in Table~\ref{tab:par}.
Based on them, the program generated the artificial light curves.
The artificial light curves have similar PSD and PDF comparing with them derived from the original light curve.
Using the simulation code, we generated $10^6$ artificial light curves for g-band data.
The 4$\sigma$ significance curve built based on these light curves is shown in left panel
of Figure~\ref{fig:gr-spec} with a green dashed line.
And the significance level of the QPO signal in the g-band is $\sim4.4\sigma$
(with $p$-value of $1\times10^{-5}$).

The same process was carried out for r-band data, and the results are shown
in right panel of Figure~\ref{fig:gr-spec}.
A power peak presents in its power spectra with a similar period as that detected in g-band.
The sine parameters and uncertainties are 0.334$\pm$0.016, 133.9$\pm$0.6~day, 58196.1$\pm$1.1, and 19.58$\pm$0.01
corresponding to $A$, $P$, $t_0$ and $A_0$, respectively. We show it
in B panel of Figure~\ref{fig:lc} with a blue dashed-dotted sine and the 4$\sigma$
significance curve in right panel of Figure~\ref{fig:gr-spec} with a green dashed line.
Its significance of the periodic signal is $\sim4.1\sigma$ (with $p$-value of $4.5\times10^{-5}$).

Considering the observation cadence performed by ZTF for the target, in our analysis,
we searched the QPO signal in a scale of frequencies between $f_{\rm max}$ = 1/7 and
$f_{\rm min}$ = 1/1350 day$^{-1}$ by employed the LSP method, where 1350 day is approximately
the total length of the data. In order to evaluate the \emph{precise} significance for the QPO signal,
we take a trial factor into our significance level calculation. The false-alarm probability (FAP) is derived
by a form of FAP=1-(1-$p$)$^N$, where $N=192$ is the number of independent frequencies in
the frequency range. The trial number is calculated by a form of $N=\frac{f_{\rm max}-f_{\rm min}}{\delta f}$,
where $\delta f$ is determined by the mean total length of the g- and r-band data (i.e. the frequency resolution).
After considering that the trial factor in the periodic signal searching,
the significance of the QPO in the g-band is 3.1$\sigma$, while in the r-band, it is 2.6$\sigma$.

Then a independent method of WWZ was used to check the results obtained by the LSP.
And a time–frequency WWZ power spectra were obtained for g- and r-band light curves,
the suitable WWZ power values were calculated by changing the parameters of wavelets to match the LSP results.
In g- and r-band data, a QPO signal presents in time–frequency power spectra over the whole-data time range,
we show them in Figure \ref{fig:gr-spec} by scaled the power values with a colorbar.
A time-averaged power spectra along the time-dimensionality were also created for WWZ power spectra,
we show them in Figure \ref{fig:gr-spec} with a black curves, and a QPO signal
presents in them too, which has the same shape and period as the LSP power spectra.

For the i-band light curve, a weak peak was also seen in its power spectra with a frequency corresponding
to the periods determined from the g- and r-band light curves.
It is worth mentioning that the i-band light curve comprises only 133 data points,
which are concentrated in three time intervals with large gaps between them, namely 2018 June,
2019 May, and 2020 May (see C panel in Figure~\ref{fig:lc}).
It is possible that the properties of QPO variabilities may not be fully captured by i-band light curve
because of these intervals. These causes may result in the QPO signal not being statistically significant
in i-band.

\begin{figure*}
\centering
\includegraphics[angle=0,scale=0.49]{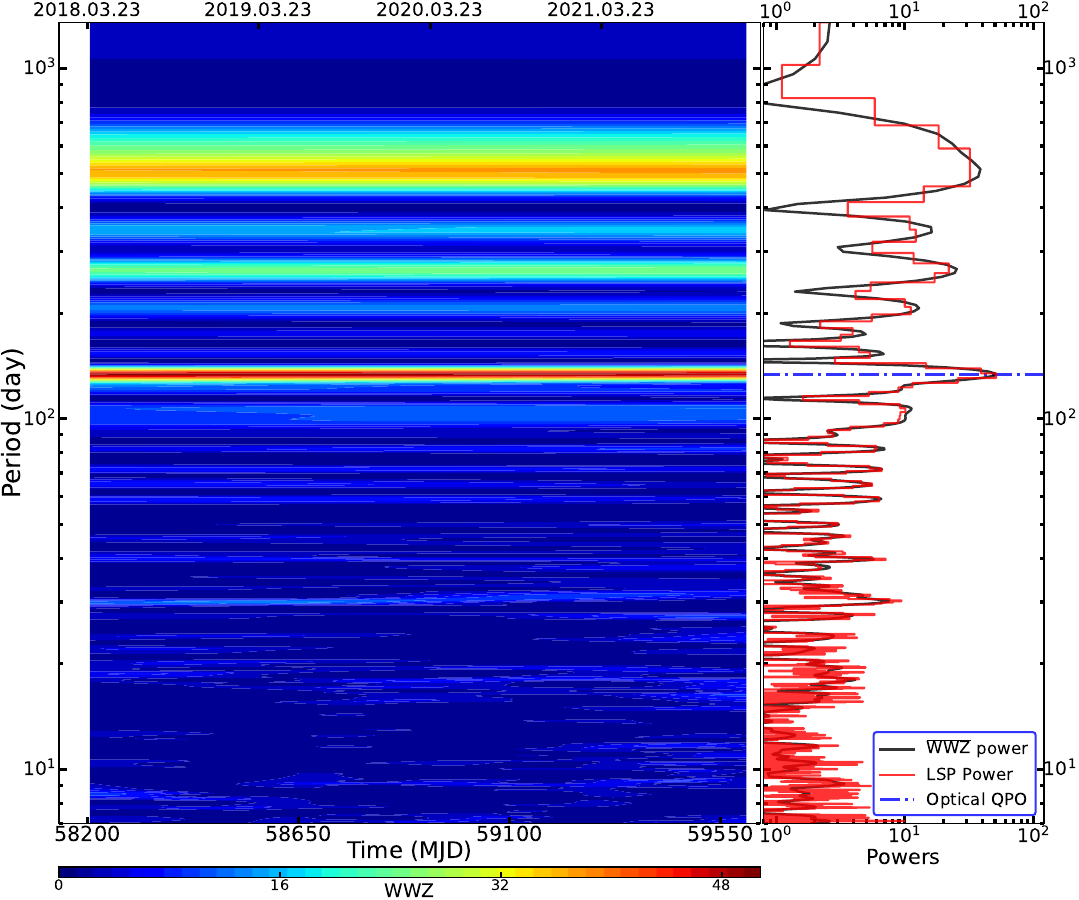}
\includegraphics[angle=0,scale=0.49]{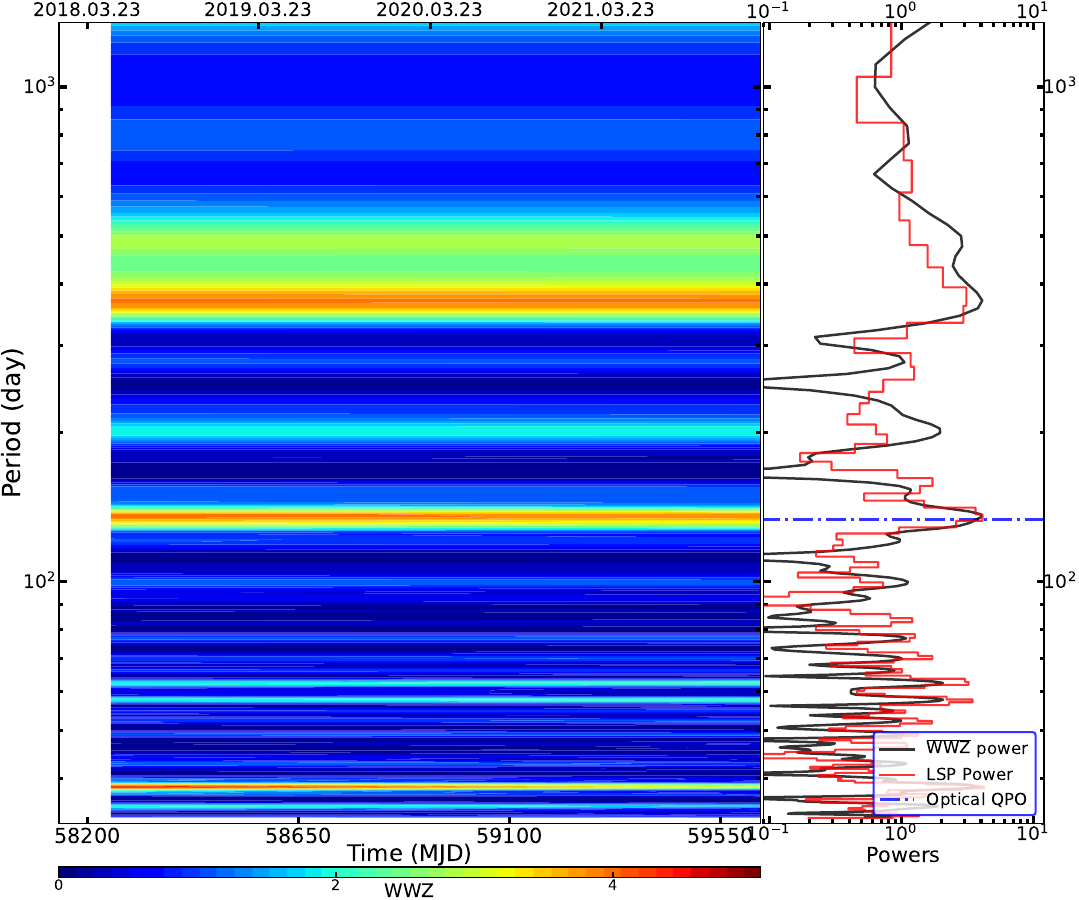}
\caption{Power spectra of TXS1206+549 constructed with the o-band and $\gamma$-ray light curves
         with the simultaneous time range of ZTF observations. A strong and weak QPO signals at
         $\sim$134~day are shown in ATLAS and \emph{Fermi}-LAT observations.
         We use a dashed-dotted line to stand for the QPO signal that detected in ZTF light curves.}
\label{fig:otherpower}
\end{figure*}

By the same procedure, we searched the periodicity in the o-band light curve from
the whole ATLAS observations, and no significant power peak in the power spectra was found.
When we focused on the simultaneous light curve as the time scale of ZTF observations,
a potential QPO signal with a similar period as that found in ZTF reveals in power spectra.
We show them in left panel of Figure \ref{fig:otherpower}.
The same case was also shown in the gamma-ray light curve, when we focuses on the simultaneous time
scale of ZTF, a weak periodic signal was detected at similar period.
We show the $\gamma$-ray results in right panel of Figure \ref{fig:otherpower}.
For comparison, we show the QPO found in ZTF data with a blue dashed-dotted line
in Figure \ref{fig:otherpower}.

\subsection{Phase-resolved Results}

\begin{figure*}
\centering
\includegraphics[angle=0,scale=0.9]{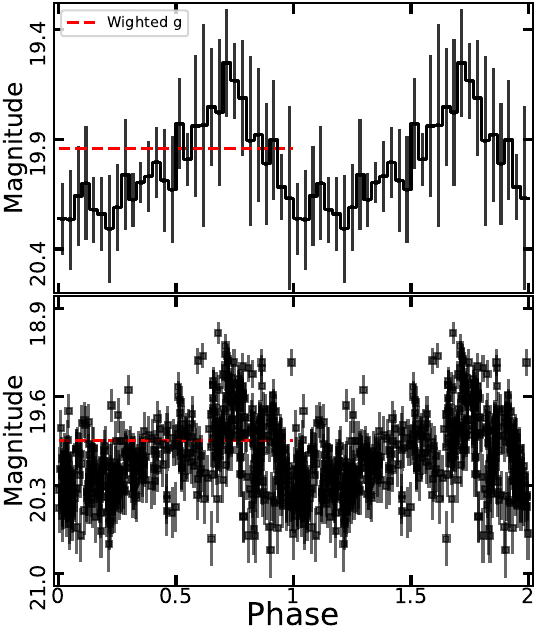}
\includegraphics[angle=0,scale=0.9]{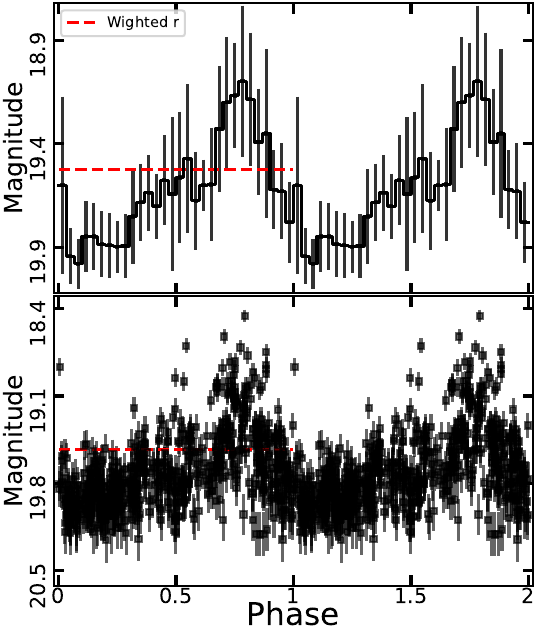}
\caption{Pulse profiles of the optical light curves in ZTF g- and r-bands.
         We folded the data points with the period cycles of 134.82 and 133.94 day
         with phase zero at MJD 58200 for g- and r-bands, respectively.
         Two cycles are plotted for clarity.
         The red dashed lines represent the weighted magnitudes by the data point errors.
         Lower panel: Data points were folded with the corresponding periods of g- and r-bands.
         Upper panel: The folded data points were averaged into 30 phase bins in one cycle,
         their error-bars were calculated by the weighted standard deviations of the magnitudes and errors.
         }
\label{fig:folded}
\end{figure*}

Based on the results of the periodic analysis, we folded the light curves with the respective periods 
that detected in g- and r-bands. The folded light curves have a same phase zero at MJD 58200.0,
which corresponding to the start time of ZTF observations. We show the folded data points
with the raw data in lower panel of Figure~\ref{fig:folded}.
Based on the lower panel, we defined 30 phase bins to average the weighted magnitudes
for each data point which falls into the corresponding phase bin. The weighted magnitudes and
their errors were calculated with the same procedure presented in Section~\ref{sec:ztf}.
In lower panel, for g-band (r-band), the minimum, maximum, and mean values are $\sim$19.55(19.10),
20.31(19.98), and 20.01(19.61) Mag, respectively.
From the folded light curves, we can see that a near-sinusoidal profile show in it.

\section{Summary and Discussion}
\label{sec:dis}

TXS 1206+549 is a radio and $\gamma$-ray loud AGN classed as NLSy1 galaxy at redshift of $z$=1.344 \citep{rss+21}.
Here, we report an optical QPO signal with a period of $\sim$134~day detected in the g- and r-band
light curves from the ZTF observations spanning approximately 4~yr, the signal presents about 10 cycles.
The QPO behavior is depicted in Figure~\ref{fig:lc} using a sinusoidal function.
We observe that some data points do not align with the periodic model.
This leads us to suspect that the target's variability is not solely characterized by the QPO behavior,
it also encompasses variations originating from other electromagnetic radiation mechanisms
across the entire frequency spectrum,
while the periodic model only represents the QPO signal at one certain single frequency.
This may be the reason of the deviations between data points and the periodic model.
After considering that the trial factor in the timing analysis,
the significance of the QPO in the g-band is 3.1$\sigma$, while in the r-band, it is 2.6$\sigma$.
Interestingly, by the same procedure, a potential QPO signal with a similar period as that
detected in ZTF was found in the simultaneous light curve of o-band from the ATLAS observations,
and a weak signal also presented in the nearly simultaneous light curve in \emph{Fermi}-LAT observations.
Based on the period detected by ZTF observations, we folded the light curves of g- and r-bands,
and a near-sinusoidal pulse profile show in them (see Figure~\ref{fig:folded}).

Recently many periodicity phenomena have been reported in the electromagnetic emissions from AGNs. 
And many mechanisms causing these periodicities are also proposed \citep[][and references therein]{1553+15},
while they are still poorly understood.
The main kinds of models for explaining these QPOs generally including
(i) jet precession/rotation or helical structure, (ii) pulsational accretion flow instabilities,
and (iii) binary SMBH system.

TXS 1206+549 is a $\gamma$-ray emitting NLSy1 galaxy \citep{rss+21}
and displays a large amplitude variabilities in $\gamma$-rays (as shown in Figure~\ref{fig:latlc}),
which make it to be one of $\gamma$-ray blazar-like sources, they generally host powerful relativistic jets.
And indeed a probable kpc-scale jet reveals in the Very Large Array Sky Survey (VLASS) data \citep{psj22}.
If TXS 1206+549's jet is precessing, or an internal helical configuration presents in its jet,
the QPO variations could arise from a varying Doppler boosting for
a periodically changing viewing angle \citep{r04,kz16}.
In the optical light curves, TXS 1206+549 has a observed period $T_{\rm obs}\sim$ 4~month,
and the intrinsic periods $T_{\rm int}$ is calculated by $T_{\rm int}=T_{\rm obs}/(1+z)\sim$2~month,
this period is in agreement with that a monthly QPO found in PKS 2247--131 by \citet{zwc+18},
they also claimed that the QPO can be explained by the model of a helical structure in the jet.

For the scenario of binary SMBHs, we can estimate the parameters of the binary system, e.g.
the separation ($D$) between the two SMBHs and the merging timescale ($t_{\rm m}$) of the two SMBH.
As the equation (26) shown in \citet{sss17}, the separation between the two SMBH can be estimated
by
\begin{equation}
D=1.3\times10^{16}(\frac{1+q}{q})\frac{T_{\rm int}}{2~\rm yr}~\rm cm,
\label{eq:sep}
\end{equation}
where $q$ stands for the mass ratio of the secondary SMBH to the primary one.
Using $T_{\rm int}\sim0.16$ (134 day) and $q\sim1$, we obtain $D\sim0.0006$~pc.
The separation of binary SMBH system might be too small.
Owing to emission of gravitational waves, the two SMBHs will merge at the timescale of
\begin{equation}
t_{\rm m}=3.9\times10^{4}q(\frac{q}{1+q})^3\frac{T_{\rm int}}{2~\rm yr}~\rm yr.
\label{eq:tm}
\end{equation}
We obtain $t_{\rm m}\sim380$ yr. This value is slightly early than a normal timescale
for the binary SMBHs merging \citep{p64,y02}.
In addition, only one radio core in TXS 1206+549 was detected in the radio VLASS observations \citep{psj22}.
In consideration of above factors, we suspect that the scenario of binary SMBHs may not be suitable
for this QPO case reported here.

We know that the NLSy1 galaxies generally have high accretion rates.
The model of pulsational accretion flow instabilities possibly induces quasi-periodic variabilities
in jet emissions. While they are generally believed to result in the short periods \citep{hmk92} at a scale of
\begin{equation}
T_{\rm int}\sim1\times\frac{M_{\rm SMBH}}{10^8~M_{\odot}}~\rm day.
\label{eq:tp}
\end{equation}
Using the mass of SMBH of a value of $1.2\times10^{8}~M_{\odot}$ for TXS 1206+549 reported in \citet{rss+21},
we obtained $T_{\rm int}\sim$1.2 day, this value is significantly lower than the period of QPO signal found here.
Therefore, this scenario may be not suitable for the QPO reported here.

It should be kept in mind that many other scenarios also have been proposed to explain the QPOs detected
in AGNs \citep{kz16}, more researches of the searchings for the multifrequency QPOs should be encouraged
to probe the underlying physical mechanism of periodicity phenomena in AGNs.

\section*{acknowledgments}
We thank anonymous referees for their comments that helped improving this work.
This work is supported in part by the National Natural Science Foundation of China
No.~12163006 and 12233006, the Basic Research Program of Yunnan Province
No.~202201AT070137, and the joint foundation of Department of Science and
Technology of Yunnan Province and Yunnan University No.~202201BF070001-020.

\bibliographystyle{aasjournal}
\bibliography{aas}

\begin{thebibliography}{}
\expandafter\ifx\csname natexlab\endcsname\relax\def\natexlab#1{#1}\fi
\providecommand{\url}[1]{\href{#1}{#1}}
\providecommand{\dodoi}[1]{doi:~\href{http://doi.org/#1}{\nolinkurl{#1}}}
\providecommand{\doeprint}[1]{\href{http://ascl.net/#1}{\nolinkurl{http://ascl.net/#1}}}
\providecommand{\doarXiv}[1]{\href{https://arxiv.org/abs/#1}{\nolinkurl{https://arxiv.org/abs/#1}}}

\bibitem[{{Ackermann} {et~al.}(2015){Ackermann}, {Ajello}, {Albert}, {Atwood},
  {Baldini}, {Ballet}, {Barbiellini}, {Bastieri}, {Becerra Gonzalez},
  {Bellazzini}, {Bissaldi}, {Blandford}, {Bloom}, {Bonino}, {Bottacini},
  {Bregeon}, {Bruel}, {Buehler}, {Buson}, {Caliandro}, {Cameron}, {Caputo},
  {Caragiulo}, {Caraveo}, {Cavazzuti}, {Cecchi}, {Chekhtman}, {Chiang},
  {Chiaro}, {Ciprini}, {Cohen-Tanugi}, {Conrad}, {Cutini}, {D'Ammando}, {de
  Angelis}, {de Palma}, {Desiante}, {Di Venere}, {Dom{\'\i}nguez}, {Drell},
  {Favuzzi}, {Fegan}, {Ferrara}, {Focke}, {Fuhrmann}, {Fukazawa}, {Fusco},
  {Gargano}, {Gasparrini}, {Giglietto}, {Giommi}, {Giordano}, {Giroletti},
  {Godfrey}, {Green}, {Grenier}, {Grove}, {Guiriec}, {Harding}, {Hays},
  {Hewitt}, {Hill}, {Horan}, {Jogler}, {J{\'o}hannesson}, {Johnson}, {Kamae},
  {Kuss}, {Larsson}, {Latronico}, {Li}, {Li}, {Longo}, {Loparco}, {Lott},
  {Lovellette}, {Lubrano}, {Magill}, {Maldera}, {Manfreda}, {Max-Moerbeck},
  {Mayer}, {Mazziotta}, {McEnery}, {Michelson}, {Mizuno}, {Monzani},
  {Morselli}, {Moskalenko}, {Murgia}, {Nuss}, {Ohno}, {Ohsugi}, {Ojha},
  {Omodei}, {Orlando}, {Ormes}, {Paneque}, {Pearson}, {Perkins}, {Perri},
  {Pesce-Rollins}, {Petrosian}, {Piron}, {Pivato}, {Porter}, {Rain{\`o}},
  {Rando}, {Razzano}, {Readhead}, {Reimer}, {Reimer}, {Schulz}, {Sgr{\`o}},
  {Siskind}, {Spada}, {Spandre}, {Spinelli}, {Suson}, {Takahashi}, {Thayer},
  {Thompson}, {Tibaldo}, {Torres}, {Tosti}, {Troja}, {Uchiyama}, {Vianello},
  {Wood}, {Wood}, {Zimmer}, {Berdyugin}, {Corbet}, {Hovatta}, {Lindfors},
  {Nilsson}, {Reinthal}, {Sillanp{\"a}{\"a}}, {Stamerra}, {Takalo}, \&
  {Valtonen}}]{1553+15}
{Ackermann}, M., {Ajello}, M., {Albert}, A., {et~al.} 2015, \apjl, 813, L41,
  \dodoi{10.1088/2041-8205/813/2/L41}

\bibitem[{{Atwood} {et~al.}(2009){Atwood}, {Abdo}, {Ackermann}, {Althouse},
  {Anderson}, {Axelsson}, {Baldini}, {Ballet}, {Band}, {Barbiellini},
  {Bartelt}, {Bastieri}, {Baughman}, {Bechtol}, {B{\'e}d{\'e}r{\`e}de},
  {Bellardi}, {Bellazzini}, {Berenji}, {Bignami}, {Bisello}, {Bissaldi},
  {Blandford}, {Bloom}, {Bogart}, {Bonamente}, {Bonnell}, {Borgland},
  {Bouvier}, {Bregeon}, {Brez}, {Brigida}, {Bruel}, {Burnett}, {Busetto},
  {Caliandro}, {Cameron}, {Caraveo}, {Carius}, {Carlson}, {Casandjian},
  {Cavazzuti}, {Ceccanti}, {Cecchi}, {Charles}, {Chekhtman}, {Cheung},
  {Chiang}, {Chipaux}, {Cillis}, {Ciprini}, {Claus}, {Cohen-Tanugi},
  {Condamoor}, {Conrad}, {Corbet}, {Corucci}, {Costamante}, {Cutini}, {Davis},
  {Decotigny}, {DeKlotz}, {Dermer}, {de Angelis}, {Digel}, {do Couto e Silva},
  {Drell}, {Dubois}, {Dumora}, {Edmonds}, {Fabiani}, {Farnier}, {Favuzzi},
  {Flath}, {Fleury}, {Focke}, {Funk}, {Fusco}, {Gargano}, {Gasparrini},
  {Gehrels}, {Gentit}, {Germani}, {Giebels}, {Giglietto}, {Giommi}, {Giordano},
  {Glanzman}, {Godfrey}, {Grenier}, {Grondin}, {Grove}, {Guillemot}, {Guiriec},
  {Haller}, {Harding}, {Hart}, {Hays}, {Healey}, {Hirayama}, {Hjalmarsdotter},
  {Horn}, {Hughes}, {J{\'o}hannesson}, {Johansson}, {Johnson}, {Johnson},
  {Johnson}, {Johnson}, {Kamae}, {Katagiri}, {Kataoka}, {Kavelaars}, {Kawai},
  {Kelly}, {Kerr}, {Klamra}, {Kn{\"o}dlseder}, {Kocian}, {Komin}, {Kuehn},
  {Kuss}, {Landriu}, {Latronico}, {Lee}, {Lee}, {Lemoine-Goumard}, {Lionetto},
  {Longo}, {Loparco}, {Lott}, {Lovellette}, {Lubrano}, {Madejski}, {Makeev},
  {Marangelli}, {Massai}, {Mazziotta}, {McEnery}, {Menon}, {Meurer},
  {Michelson}, {Minuti}, {Mirizzi}, {Mitthumsiri}, {Mizuno}, {Moiseev},
  {Monte}, {Monzani}, {Moretti}, {Morselli}, {Moskalenko}, {Murgia},
  {Nakamori}, {Nishino}, {Nolan}, {Norris}, {Nuss}, {Ohno}, {Ohsugi}, {Omodei},
  {Orlando}, {Ormes}, {Paccagnella}, {Paneque}, {Panetta}, {Parent}, {Pearce},
  {Pepe}, {Perazzo}, {Pesce-Rollins}, {Picozza}, {Pieri}, {Pinchera}, {Piron},
  {Porter}, {Poupard}, {Rain{\`o}}, {Rando}, {Rapposelli}, {Razzano}, {Reimer},
  {Reimer}, {Reposeur}, {Reyes}, {Ritz}, {Rochester}, {Rodriguez}, {Romani},
  {Roth}, {Russell}, {Ryde}, {Sabatini}, {Sadrozinski}, {Sanchez}, {Sander},
  {Sapozhnikov}, {Parkinson}, {Scargle}, {Schalk}, {Scolieri}, {Sgr{\`o}},
  {Share}, {Shaw}, {Shimokawabe}, {Shrader}, {Sierpowska-Bartosik}, {Siskind},
  {Smith}, {Smith}, {Spandre}, {Spinelli}, {Starck}, {Stephens}, {Strickman},
  {Strong}, {Suson}, {Tajima}, {Takahashi}, {Takahashi}, {Tanaka}, {Tenze},
  {Tether}, {Thayer}, {Thayer}, {Thompson}, {Tibaldo}, {Tibolla}, {Torres},
  {Tosti}, {Tramacere}, {Turri}, {Usher}, {Vilchez}, {Vitale}, {Wang},
  {Watters}, {Winer}, {Wood}, {Ylinen}, \& {Ziegler}}]{Atwood2009}
{Atwood}, W.~B., {Abdo}, A.~A., {Ackermann}, M., {et~al.} 2009, \apj, 697,
  1071, \dodoi{10.1088/0004-637X/697/2/1071}

\bibitem[{{Banerjee} {et~al.}(2022){Banerjee}, {Negi}, {Joshi}, {Kumar},
  {Wiita}, {Chand}, {Rawat}, {Wu}, \& {Ho}}]{bnj+22}
{Banerjee}, A., {Negi}, V., {Joshi}, R., {et~al.} 2022, arXiv e-prints,
  arXiv:2210.07266, \dodoi{10.48550/arXiv.2210.07266}

\bibitem[{{Barret} \& {Vaughan}(2012)}]{bv12}
{Barret}, D., \& {Vaughan}, S. 2012, \apj, 746, 131,
  \dodoi{10.1088/0004-637X/746/2/131}

\bibitem[{{Bellm} {et~al.}(2019){Bellm}, {Kulkarni}, {Graham}, {Dekany},
  {Smith}, {Riddle}, {Masci}, {Helou}, {Prince}, {Adams}, {Barbarino},
  {Barlow}, {Bauer}, {Beck}, {Belicki}, {Biswas}, {Blagorodnova}, {Bodewits},
  {Bolin}, {Brinnel}, {Brooke}, {Bue}, {Bulla}, {Burruss}, {Cenko}, {Chang},
  {Connolly}, {Coughlin}, {Cromer}, {Cunningham}, {De}, {Delacroix}, {Desai},
  {Duev}, {Eadie}, {Farnham}, {Feeney}, {Feindt}, {Flynn}, {Franckowiak},
  {Frederick}, {Fremling}, {Gal-Yam}, {Gezari}, {Giomi}, {Goldstein},
  {Golkhou}, {Goobar}, {Groom}, {Hacopians}, {Hale}, {Henning}, {Ho}, {Hover},
  {Howell}, {Hung}, {Huppenkothen}, {Imel}, {Ip}, {Ivezi{\'c}}, {Jackson},
  {Jones}, {Juric}, {Kasliwal}, {Kaspi}, {Kaye}, {Kelley}, {Kowalski},
  {Kramer}, {Kupfer}, {Landry}, {Laher}, {Lee}, {Lin}, {Lin}, {Lunnan},
  {Giomi}, {Mahabal}, {Mao}, {Miller}, {Monkewitz}, {Murphy}, {Ngeow},
  {Nordin}, {Nugent}, {Ofek}, {Patterson}, {Penprase}, {Porter}, {Rauch},
  {Rebbapragada}, {Reiley}, {Rigault}, {Rodriguez}, {van Roestel}, {Rusholme},
  {van Santen}, {Schulze}, {Shupe}, {Singer}, {Soumagnac}, {Stein}, {Surace},
  {Sollerman}, {Szkody}, {Taddia}, {Terek}, {Van Sistine}, {van Velzen},
  {Vestrand}, {Walters}, {Ward}, {Ye}, {Yu}, {Yan}, \& {Zolkower}}]{ztf2+19}
{Bellm}, E.~C., {Kulkarni}, S.~R., {Graham}, M.~J., {et~al.} 2019, \pasp, 131,
  018002, \dodoi{10.1088/1538-3873/aaecbe}

\bibitem[{{Bhatta}(2017)}]{b17}
{Bhatta}, G. 2017, \apj, 847, 7, \dodoi{10.3847/1538-4357/aa86ed}

\bibitem[{{Bhatta}(2018)}]{b18}
---. 2018, Galaxies, 6, 136, \dodoi{10.3390/galaxies6040136}

\bibitem[{{Chen} {et~al.}(2022){Chen}, {Yi}, {Gong}, {Yang}, {Chen}, {Chang},
  \& {Mao}}]{cyg+22}
{Chen}, J., {Yi}, T., {Gong}, Y., {et~al.} 2022, \apj, 938, 8,
  \dodoi{10.3847/1538-4357/ac91c3}

\bibitem[{{Emmanoulopoulos} {et~al.}(2013){Emmanoulopoulos}, {McHardy}, \&
  {Papadakis}}]{emp13}
{Emmanoulopoulos}, D., {McHardy}, I.~M., \& {Papadakis}, I.~E. 2013, \mnras,
  433, 907, \dodoi{10.1093/mnras/stt764}

\bibitem[{{Fan} \& {Lin}(2000)}]{fl00}
{Fan}, J.~H., \& {Lin}, R.~G. 2000, \aap, 355, 880.
\newblock \doarXiv{astro-ph/0001028}

\bibitem[{{Fermi-LAT collaboration} {et~al.}(2022){Fermi-LAT collaboration},
  {:}, {Abdollahi}, {Acero}, {Baldini}, {Ballet}, {Bastieri}, {Bellazzini},
  {Berenji}, {Berretta}, {Bissaldi}, {Blandford}, {Bloom}, {Bonino}, {Brill},
  {Britto}, {Bruel}, {Burnett}, {Buson}, {Cameron}, {Caputo}, {Caraveo},
  {Castro}, {Chaty}, {Cheung}, {Chiaro}, {Cibrario}, {Ciprini},
  {Coronado-Blazquez}, {Crnogorcevic}, {Cutini}, {D'Ammando}, {De Gaetano},
  {Digel}, {Di Lalla}, {Dirirsa}, {Di Venere}, {Dominguez}, {Fallah Ramazani},
  {Fegan}, {Ferrara}, {Fiori}, {Fleischhack}, {Franckowiak}, {Fukazawa},
  {Funk}, {Fusco}, {Galanti}, {Gammaldi}, {Gargano}, {Garrappa}, {Gasparrini},
  {Giacchino}, {Giglietto}, {Giordano}, {Giroletti}, {Glanzman}, {Green},
  {Grenier}, {Grondin}, {Guillemot}, {Guiriec}, {Gustafsson}, {Harding},
  {Hays}, {Hewitt}, {Horan}, {Hou}, {Johannesson}, {Karwin}, {Kayanoki},
  {Kerr}, {Kuss}, {Landriu}, {Larsson}, {Latronico}, {Lemoine-Goumard}, {Li},
  {Liodakis}, {Longo}, {Loparco}, {Lott}, {Lubrano}, {Maldera}, {Malyshev},
  {Manfreda}, {Marti-Devesa}, {Mazziotta}, {Mereu}, {Meyer}, {Michelson},
  {Mirabal}, {Mitthumsiri}, {Mizuno}, {Moiseev}, {Monzani}, {Morselli},
  {Moskalenko}, {Negro}, {Nuss}, {Omodei}, {Orienti}, {Orlando}, {Paneque},
  {Pei}, {Perkins}, {Persic}, {Pesce-Rollins}, {Petrosian}, {Pillera}, {Poon},
  {Porter}, {Principe}, {Raino}, {Rando}, {Rani}, {Razzano}, {Razzaque},
  {Reimer}, {Reimer}, {Reposeur}, {Sanchez-Conde}, {Saz Parkinson}, {Scotton},
  {Serini}, {Sgro}, {Siskind}, {Smith}, {Spandre}, {Spinelli}, {Sueoka},
  {Suson}, {Tajima}, {Tak}, {Thayer}, {Thompson}, {Torres}, {Troja},
  {Valverde}, {Wood}, \& {Zaharijas}}]{4fgl-dr3}
{Fermi-LAT collaboration}, {:}, {Abdollahi}, S., {et~al.} 2022, arXiv e-prints,
  arXiv:2201.11184.
\newblock \doarXiv{2201.11184}

\bibitem[{{Foster}(1996)}]{f96}
{Foster}, G. 1996, \aj, 112, 1709, \dodoi{10.1086/118137}

\bibitem[{{Gierli{\'n}ski} {et~al.}(2008){Gierli{\'n}ski}, {Middleton}, {Ward},
  \& {Done}}]{gmw+08}
{Gierli{\'n}ski}, M., {Middleton}, M., {Ward}, M., \& {Done}, C. 2008, \nat,
  455, 369, \dodoi{10.1038/nature07277}

\bibitem[{{Gong} {et~al.}(2023){Gong}, {Tian}, {Zhou}, {Yi}, \&
  {Fang}}]{gtz+23}
{Gong}, Y., {Tian}, S., {Zhou}, L., {Yi}, T., \& {Fang}, J. 2023, arXiv
  e-prints, arXiv:2304.03085, \dodoi{10.48550/arXiv.2304.03085}

\bibitem[{{Gong} {et~al.}(2022){Gong}, {Zhou}, {Yuan}, {Zhang}, {Yi}, \&
  {Fang}}]{gzy+22}
{Gong}, Y., {Zhou}, L., {Yuan}, M., {et~al.} 2022, \apj, 931, 168,
  \dodoi{10.3847/1538-4357/ac6c8c}

\bibitem[{{Gonz{\'a}lez-Mart{\'\i}n} \& {Vaughan}(2012)}]{gv12}
{Gonz{\'a}lez-Mart{\'\i}n}, O., \& {Vaughan}, S. 2012, \aap, 544, A80,
  \dodoi{10.1051/0004-6361/201219008}

\bibitem[{{Goodrich}(1989)}]{g89}
{Goodrich}, R.~W. 1989, \apj, 342, 224, \dodoi{10.1086/167586}

\bibitem[{{Graham} {et~al.}(2019){Graham}, {Kulkarni}, {Bellm}, {Adams},
  {Barbarino}, {Blagorodnova}, {Bodewits}, {Bolin}, {Brady}, {Cenko}, {Chang},
  {Coughlin}, {De}, {Eadie}, {Farnham}, {Feindt}, {Franckowiak}, {Fremling},
  {Gezari}, {Ghosh}, {Goldstein}, {Golkhou}, {Goobar}, {Ho}, {Huppenkothen},
  {Ivezi{\'c}}, {Jones}, {Juric}, {Kaplan}, {Kasliwal}, {Kelley}, {Kupfer},
  {Lee}, {Lin}, {Lunnan}, {Mahabal}, {Miller}, {Ngeow}, {Nugent}, {Ofek},
  {Prince}, {Rauch}, {van Roestel}, {Schulze}, {Singer}, {Sollerman}, {Taddia},
  {Yan}, {Ye}, {Yu}, {Barlow}, {Bauer}, {Beck}, {Belicki}, {Biswas}, {Brinnel},
  {Brooke}, {Bue}, {Bulla}, {Burruss}, {Connolly}, {Cromer}, {Cunningham},
  {Dekany}, {Delacroix}, {Desai}, {Duev}, {Feeney}, {Flynn}, {Frederick},
  {Gal-Yam}, {Giomi}, {Groom}, {Hacopians}, {Hale}, {Helou}, {Henning},
  {Hover}, {Hillenbrand}, {Howell}, {Hung}, {Imel}, {Ip}, {Jackson}, {Kaspi},
  {Kaye}, {Kowalski}, {Kramer}, {Kuhn}, {Landry}, {Laher}, {Mao}, {Masci},
  {Monkewitz}, {Murphy}, {Nordin}, {Patterson}, {Penprase}, {Porter},
  {Rebbapragada}, {Reiley}, {Riddle}, {Rigault}, {Rodriguez}, {Rusholme}, {van
  Santen}, {Shupe}, {Smith}, {Soumagnac}, {Stein}, {Surace}, {Szkody}, {Terek},
  {Van Sistine}, {van Velzen}, {Vestrand}, {Walters}, {Ward}, {Zhang}, \&
  {Zolkower}}]{ztf1+19}
{Graham}, M.~J., {Kulkarni}, S.~R., {Bellm}, E.~C., {et~al.} 2019, \pasp, 131,
  078001, \dodoi{10.1088/1538-3873/ab006c}

\bibitem[{{Honma} {et~al.}(1992){Honma}, {Matsumoto}, \& {Kato}}]{hmk92}
{Honma}, F., {Matsumoto}, R., \& {Kato}, S. 1992, \pasj, 44, 529

\bibitem[{{Kidger} {et~al.}(1992){Kidger}, {Takalo}, \& {Sillanpaa}}]{kts92}
{Kidger}, M., {Takalo}, L., \& {Sillanpaa}, A. 1992, \aap, 264, 32

\bibitem[{{King} {et~al.}(2013){King}, {Hovatta}, {Max-Moerbeck}, {Meier},
  {Pearson}, {Readhead}, {Reeves}, {Richards}, \& {Shepherd}}]{khm+13}
{King}, O.~G., {Hovatta}, T., {Max-Moerbeck}, W., {et~al.} 2013, \mnras, 436,
  L114, \dodoi{10.1093/mnrasl/slt125}

\bibitem[{{Komossa} \& {Zensus}(2016)}]{kz16}
{Komossa}, S., \& {Zensus}, J.~A. 2016, in Star Clusters and Black Holes in
  Galaxies across Cosmic Time, ed. Y.~{Meiron}, S.~{Li}, F.~K. {Liu}, \&
  R.~{Spurzem}, Vol. 312, 13--25, \dodoi{10.1017/S1743921315007395}

\bibitem[{{Li} {et~al.}(2017){Li}, {Luo}, {Yang}, {Yang}, {Cai}, \&
  {Yang}}]{lly+17}
{Li}, X.-P., {Luo}, Y.-H., {Yang}, H.-Y., {et~al.} 2017, \apj, 847, 8,
  \dodoi{10.3847/1538-4357/aa86ee}

\bibitem[{{Lin} {et~al.}(2013){Lin}, {Irwin}, {Godet}, {Webb}, \&
  {Barret}}]{lig+13}
{Lin}, D., {Irwin}, J.~A., {Godet}, O., {Webb}, N.~A., \& {Barret}, D. 2013,
  \apjl, 776, L10, \dodoi{10.1088/2041-8205/776/1/L10}

\bibitem[{{Lomb}(1976)}]{l76}
{Lomb}, N.~R. 1976, \apss, 39, 447, \dodoi{10.1007/BF00648343}

\bibitem[{{Mineshige} {et~al.}(2000){Mineshige}, {Kawaguchi}, {Takeuchi}, \&
  {Hayashida}}]{min+00}
{Mineshige}, S., {Kawaguchi}, T., {Takeuchi}, M., \& {Hayashida}, K. 2000,
  \pasj, 52, 499, \dodoi{10.1093/pasj/52.3.499}

\bibitem[{{Ojha} {et~al.}(2020){Ojha}, {Chand}, {Dewangan}, \&
  {Rakshit}}]{ojh+20}
{Ojha}, V., {Chand}, H., {Dewangan}, G.~C., \& {Rakshit}, S. 2020, \apj, 896,
  95, \dodoi{10.3847/1538-4357/ab94ac}

\bibitem[{{Osterbrock} \& {Pogge}(1985)}]{op85}
{Osterbrock}, D.~E., \& {Pogge}, R.~W. 1985, \apj, 297, 166,
  \dodoi{10.1086/163513}

\bibitem[{{Otero-Santos} {et~al.}(2020){Otero-Santos}, {Acosta-Pulido},
  {Becerra Gonz{\'a}lez}, {Raiteri}, {Larionov}, {Pe{\~n}il}, {Smith},
  {Ballester Niebla}, {Borman}, {Carnerero}, {Castro Segura}, {Grishina},
  {Kopatskaya}, {Larionova}, {Morozova}, {Nikiforova}, {Savchenko},
  {Troitskaya}, {Troitsky}, {Vasilyev}, \& {Villata}}]{oab+20}
{Otero-Santos}, J., {Acosta-Pulido}, J.~A., {Becerra Gonz{\'a}lez}, J.,
  {et~al.} 2020, \mnras, 492, 5524, \dodoi{10.1093/mnras/staa134}

\bibitem[{{Pajdosz-{\'S}mierciak} {et~al.}(2022){Pajdosz-{\'S}mierciak},
  {{\'S}mierciak}, \& {Jamrozy}}]{psj22}
{Pajdosz-{\'S}mierciak}, U., {{\'S}mierciak}, B., \& {Jamrozy}, M. 2022,
  \mnras, 514, 2122, \dodoi{10.1093/mnras/stac1372}

\bibitem[{{Pan} {et~al.}(2016){Pan}, {Yuan}, {Yao}, {Zhou}, {Liu}, {Zhou}, \&
  {Zhang}}]{pyy+16}
{Pan}, H.-W., {Yuan}, W., {Yao}, S., {et~al.} 2016, \apjl, 819, L19,
  \dodoi{10.3847/2041-8205/819/2/L19}

\bibitem[{{Peters}(1964)}]{p64}
{Peters}, P.~C. 1964, Physical Review, 136, 1224,
  \dodoi{10.1103/PhysRev.136.B1224}

\bibitem[{{Rakshit} {et~al.}(2021){Rakshit}, {Schramm}, {Stalin}, {Tanaka},
  {Paliya}, {Pal}, {Kotilainen}, \& {Shin}}]{rss+21}
{Rakshit}, S., {Schramm}, M., {Stalin}, C.~S., {et~al.} 2021, \mnras, 504, L22,
  \dodoi{10.1093/mnrasl/slab031}

\bibitem[{{Rakshit} \& {Stalin}(2017)}]{rs17}
{Rakshit}, S., \& {Stalin}, C.~S. 2017, \apj, 842, 96,
  \dodoi{10.3847/1538-4357/aa72f4}

\bibitem[{{Rakshit} {et~al.}(2017){Rakshit}, {Stalin}, {Chand}, \&
  {Zhang}}]{rak+17}
{Rakshit}, S., {Stalin}, C.~S., {Chand}, H., \& {Zhang}, X.-G. 2017, \apjs,
  229, 39, \dodoi{10.3847/1538-4365/aa6971}

\bibitem[{{Ren} {et~al.}(2021){Ren}, {Zhang}, {Zhang}, {Ding}, {Yang}, {Li},
  {Yan}, \& {Xu}}]{rzz+21}
{Ren}, G.-W., {Zhang}, H.-J., {Zhang}, X., {et~al.} 2021, Research in Astronomy
  and Astrophysics, 21, 075, \dodoi{10.1088/1674-4527/21/3/075}

\bibitem[{{Rieger}(2004)}]{r04}
{Rieger}, F.~M. 2004, \apjl, 615, L5, \dodoi{10.1086/426018}

\bibitem[{{Sandrinelli} {et~al.}(2014){Sandrinelli}, {Covino}, \&
  {Treves}}]{sct14}
{Sandrinelli}, A., {Covino}, S., \& {Treves}, A. 2014, \apjl, 793, L1,
  \dodoi{10.1088/2041-8205/793/1/L1}

\bibitem[{{Sandrinelli} {et~al.}(2016){Sandrinelli}, {Covino}, \&
  {Treves}}]{sct16}
---. 2016, \apj, 820, 20, \dodoi{10.3847/0004-637X/820/1/20}

\bibitem[{{Scargle}(1982)}]{s82}
{Scargle}, J.~D. 1982, \apj, 263, 835, \dodoi{10.1086/160554}

\bibitem[{{Shuder} \& {Osterbrock}(1981)}]{so81}
{Shuder}, J.~M., \& {Osterbrock}, D.~E. 1981, \apj, 250, 55,
  \dodoi{10.1086/159347}

\bibitem[{{Sobacchi} {et~al.}(2017){Sobacchi}, {Sormani}, \&
  {Stamerra}}]{sss17}
{Sobacchi}, E., {Sormani}, M.~C., \& {Stamerra}, A. 2017, \mnras, 465, 161,
  \dodoi{10.1093/mnras/stw2684}

\bibitem[{{Tonry} {et~al.}(2018){Tonry}, {Denneau}, {Heinze}, {Stalder},
  {Smith}, {Smartt}, {Stubbs}, {Weiland}, \& {Rest}}]{allas18}
{Tonry}, J.~L., {Denneau}, L., {Heinze}, A.~N., {et~al.} 2018, \pasp, 130,
  064505, \dodoi{10.1088/1538-3873/aabadf}

\bibitem[{{Ulrich} {et~al.}(1997){Ulrich}, {Maraschi}, \& {Urry}}]{umu97}
{Ulrich}, M.-H., {Maraschi}, L., \& {Urry}, C.~M. 1997, \araa, 35, 445,
  \dodoi{10.1146/annurev.astro.35.1.445}

\bibitem[{{Valtonen} {et~al.}(2006){Valtonen}, {Lehto}, {Sillanp{\"a}{\"a}},
  {Nilsson}, {Mikkola}, {Hudec}, {Basta}, {Ter{\"a}sranta}, {Haque}, \&
  {Rampadarath}}]{vls+06}
{Valtonen}, M.~J., {Lehto}, H.~J., {Sillanp{\"a}{\"a}}, A., {et~al.} 2006,
  \apj, 646, 36, \dodoi{10.1086/504884}

\bibitem[{{Valtonen} {et~al.}(2008){Valtonen}, {Lehto}, {Nilsson}, {Heidt},
  {Takalo}, {Sillanp{\"a}{\"a}}, {Villforth}, {Kidger}, {Poyner}, {Pursimo},
  {Zola}, {Wu}, {Zhou}, {Sadakane}, {Drozdz}, {Koziel}, {Marchev}, {Ogloza},
  {Porowski}, {Siwak}, {Stachowski}, {Winiarski}, {Hentunen}, {Nissinen},
  {Liakos}, \& {Dogru}}]{vln+08}
{Valtonen}, M.~J., {Lehto}, H.~J., {Nilsson}, K., {et~al.} 2008, \nat, 452,
  851, \dodoi{10.1038/nature06896}

\bibitem[{{Valtonen} {et~al.}(2023){Valtonen}, {Zola}, {Gopakumar},
  {L{\"a}hteenm{\"a}ki}, {Tornikoski}, {Dey}, {Gupta}, {Pursimo}, {Knudstrup},
  {Gomez}, {Hudec}, {Jel{\'\i}nek}, {{\v{S}}trobl}, {Berdyugin}, {Ciprini},
  {Reichart}, {Kouprianov}, {Matsumoto}, {Drozdz}, {Mugrauer}, {Sadun},
  {Zejmo}, {Sillanp{\"a}{\"a}}, {Lehto}, {Nilsson}, {Imazawa}, \&
  {Uemura}}]{vzg+23}
{Valtonen}, M.~J., {Zola}, S., {Gopakumar}, A., {et~al.} 2023, \mnras,
  \dodoi{10.1093/mnras/stad922}

\bibitem[{{Yan} {et~al.}(2018){Yan}, {Zhou}, {Zhang}, {Zhu}, \&
  {Wang}}]{yzz+18}
{Yan}, D., {Zhou}, J., {Zhang}, P., {Zhu}, Q., \& {Wang}, J. 2018, \apj, 867,
  53, \dodoi{10.3847/1538-4357/aae48a}

\bibitem[{{Yang} {et~al.}(2021){Yang}, {Yan}, {Zhang}, {Dai}, \&
  {Zhang}}]{yyz+21}
{Yang}, S., {Yan}, D., {Zhang}, P., {Dai}, B., \& {Zhang}, L. 2021, \apj, 907,
  105, \dodoi{10.3847/1538-4357/abcbff}

\bibitem[{{Yu}(2002)}]{y02}
{Yu}, Q. 2002, \mnras, 331, 935, \dodoi{10.1046/j.1365-8711.2002.05242.x}

\bibitem[{{Zechmeister} \& {K{\"u}rster}(2009)}]{zk09}
{Zechmeister}, M., \& {K{\"u}rster}, M. 2009, \aap, 496, 577,
  \dodoi{10.1051/0004-6361:200811296}

\bibitem[{{Zhang} {et~al.}(2014){Zhang}, {Zhao}, {Wang}, \& {Dai}}]{zzw+14}
{Zhang}, B.-K., {Zhao}, X.-Y., {Wang}, C.-X., \& {Dai}, B.-Z. 2014, Research in
  Astronomy and Astrophysics, 14, 933, \dodoi{10.1088/1674-4527/14/8/004}

\bibitem[{{Zhang} {et~al.}(2021){Zhang}, {Yan}, {Zhang}, {Yang}, \&
  {Zhang}}]{zyz+21}
{Zhang}, H., {Yan}, D., {Zhang}, P., {Yang}, S., \& {Zhang}, L. 2021, \apj,
  919, 58, \dodoi{10.3847/1538-4357/ac0cf0}

\bibitem[{{Zhang} {et~al.}(2017{\natexlab{a}}){Zhang}, {Zhang}, {Zhu}, {Yi},
  {Yao}, {Lu}, \& {Liang}}]{zzz+17}
{Zhang}, J., {Zhang}, H.-M., {Zhu}, Y.-K., {et~al.} 2017{\natexlab{a}}, \apj,
  849, 42, \dodoi{10.3847/1538-4357/aa8b7a}

\bibitem[{{Zhang} \& {Wang}(2021)}]{zw21}
{Zhang}, P., \& {Wang}, Z. 2021, \apj, 914, 1, \dodoi{10.3847/1538-4357/abfafd}

\bibitem[{{Zhang} \& {Wang}(2022)}]{zw22}
---. 2022, \apj, 934, 3, \dodoi{10.3847/1538-4357/ac778f}

\bibitem[{{Zhang} {et~al.}(2022){Zhang}, {Wang}, {Gurwell}, \&
  {Wiita}}]{zwg+22}
{Zhang}, P., {Wang}, Z., {Gurwell}, M., \& {Wiita}, P.~J. 2022, \apj, 925, 207,
  \dodoi{10.3847/1538-4357/ac425c}

\bibitem[{{Zhang} {et~al.}(2017{\natexlab{b}}){Zhang}, {Zhang}, {Yan}, {Fan},
  \& {Liu}}]{zzy+17}
{Zhang}, P., {Zhang}, P.-f., {Yan}, J.-z., {Fan}, Y.-z., \& {Liu}, Q.-z.
  2017{\natexlab{b}}, \apj, 849, 9, \dodoi{10.3847/1538-4357/aa8d6e}

\bibitem[{{Zhang} {et~al.}(2017{\natexlab{c}}){Zhang}, {Yan}, {Liao}, \&
  {Wang}}]{zyl+17}
{Zhang}, P.-f., {Yan}, D.-h., {Liao}, N.-h., \& {Wang}, J.-c.
  2017{\natexlab{c}}, \apj, 835, 260, \dodoi{10.3847/1538-4357/835/2/260}

\bibitem[{{Zhang} {et~al.}(2017{\natexlab{d}}){Zhang}, {Yan}, {Zhou}, {Fan},
  {Wang}, \& {Zhang}}]{zyz+17}
{Zhang}, P.-F., {Yan}, D.-H., {Zhou}, J.-N., {et~al.} 2017{\natexlab{d}}, \apj,
  845, 82, \dodoi{10.3847/1538-4357/aa7ecd}

\bibitem[{{Zhang} {et~al.}(2020){Zhang}, {Yan}, {Zhou}, {Wang}, \&
  {Zhang}}]{zyz+20}
{Zhang}, P.-f., {Yan}, D.-h., {Zhou}, J.-n., {Wang}, J.-c., \& {Zhang}, L.
  2020, \apj, 891, 163, \dodoi{10.3847/1538-4357/ab71fe}

\bibitem[{{Zhang} {et~al.}(2018){Zhang}, {Zhang}, {Liao}, {Yan}, {Fan}, \&
  {Liu}}]{zzl+18}
{Zhang}, P.-f., {Zhang}, P., {Liao}, N.-h., {et~al.} 2018, \apj, 853, 193,
  \dodoi{10.3847/1538-4357/aaa29a}

\bibitem[{{Zhang}(2022)}]{z22}
{Zhang}, X. 2022, \mnras, 516, 3650, \dodoi{10.1093/mnras/stac2531}

\bibitem[{{Zhou} {et~al.}(2018){Zhou}, {Wang}, {Chen}, {Wiita},
  {Vadakkumthani}, {Morrell}, {Zhang}, \& {Zhang}}]{zwc+18}
{Zhou}, J., {Wang}, Z., {Chen}, L., {et~al.} 2018, Nature Communications, 9,
  4599, \dodoi{10.1038/s41467-018-07103-2}

\end{thebibliography}
\end{document}